\documentclass[10pt,twoside]{article}
\usepackage{amssymb}
\usepackage{graphicx}
\usepackage{amsmath}
\usepackage{psfrag}

\begin{document}

\title{Flavor changing neutral currents from lepton and B decays in the two
Higgs doublet model}
\author{Rodolfo A. Diaz\thanks{%
radiazs@unal.edu.co}, R. Martinez\thanks{%
remartinezm@unal.edu.co}, and Carlos E. Sandoval\thanks{%
cesandovalu@unal.edu.co} \\
%EndAName
Universidad Nacional de Colombia, \\
Departamento de F\'{\i}sica. Bogot\'{a}, Colombia.}
\date{}
\maketitle
\vspace{-9mm}
\begin{abstract}
Constraints on the whole spectrum of lepton flavor violating vertices are
shown in the context of the standard two Higgs doublet model. The vertex
involving the $e-\tau $ mixing is much more constrained than the others, and
the decays proportional to such vertex are usually very supressed. On the
other hand, bounds on the quark sector are obtained from leptonic decays of
the $B_{d,s}^{0}$ mesons and from $\Delta M_{B_{d}^{0}}$. We emphasize that
although the $B_{d}^{0}-\overline{B}_{d}^{0}$ mixing restricts severely the $%
d-b$ mixing vertex, the upper bound for this vertex could still give a
sizeable contribution to the decay $B_{d}^{0}\rightarrow \mu \overline{\mu }$
respect to the standard model contribution, from which we see that such
vertex could still play a role in the phenomenology.

PACS: \{12.60.Fr, 12.15.Mm, 12.15.Ff, 11.30.Hv\}

Keywords: Two Higgs doublets model, flavor changing neutral currents, lepton
decays, B decays.
\end{abstract}

\vspace{-4mm}

\section{Introduction}

Many extensions of the standard model leads naturally to flavor changing
neutral currents (FCNC) in quark and lepton sectors. This is the case of
models with an extended Higgs sector. However, owing to the high suppression
imposed by experiments, several mechanisms has been used to get rid of them,
such as discrete symmetries \cite{Glashow}, permutation symmetries \cite{Lin}%
, and different textures of Yukawa couplings \cite{Zhou}. Notwithstanding,
the increasing evidence on neutrino oscillations seems to show the existence
of mass terms for neutrinos as well as of family lepton flavor violation
(LFV) \cite{Fukuda}. Such fact has inspired the study of many scenarios that
predict LFV processes as in the case of SUSY theories with R-parity broken 
\cite{R1}, SU(5) SUSY models with right-handed neutrinos \cite{Baek}, models
with heavy Majorana neutrinos \cite{Gvetic}, and multi-Higgs doublet models
with right-handed neutrinos for each lepton generation \cite{Lavoura}. On
the other hand, LFV in the charged sector has been also examined in models
such as SUSY GUT \cite{Okada}, and the two Higgs doublet model (2HDM) \cite%
{Zhou, us}.

In the charged lepton sector, searches for FCNC have been carried out
through leptonic and semileptonic decays of $K$ and $B$ mesons \cite{KB Col}%
, as well as purely leptonic processes \cite{Lepton Col}. On the other hand,
some collaborations plan to improve current upper limits of some LFV decays
by several orders of magnitude, by increasing the statistics \cite{Improve
bounds}. Other possible sources of improvement lies on the Fermilab Tevatron
and LHC by means of LFV Higgs boson decays. Further potential sources to
look for Higgs mediated FCNC lie on the muon colliders. It is because they
have the potentiality to produce Higgs bosons in the $s-$channel, with
substantial rate production at the Higgs mass resonance \cite{Reina}. From
the theoretical point of view, since Higgs Yukawa couplings are usually
proportional to the lepton mass, they give an important enhancement to cross
sections with Higgs mediated $s-$channels, respect to the ones in an $%
e^{+}e^{-}$ collider.

As for the quark sector, it is well known that the data from $K^{0}-%
\overline{K}^{0}$ and $B^{0}-\overline{B}^{0}$ mixing put severe bounds on
the flavor changing couplings involving the first family \cite{Reina, Xiao}.
Indeed, this fact was one of the motivations to implement a discrete
symmetry in the 2HDM in order to suppress FCNC effects \cite{Glashow}. This
fact in turn motivated the construction of a parameterization in the 2HDM in
which the FC vertices involving the first family are neglected, and assume
that the only non-vanishing couplings are $\lambda _{tt},\lambda _{bb}$\ 
\cite{Chao}. Based on this assumption, constraints on $\lambda _{tt}$ and $%
\lambda _{bb}$ from $B^{0}-\overline{B}^{0}$ and lower bounds on $m_{H^{\pm
}}$ from the CLEO data of $b\rightarrow s\gamma $ have been estimated \cite%
{Xiao}. In this scenario, the tree-level diagrams for $B_{d}^{0}-\overline{B}%
_{d}^{0}\ $are neglected and the box diagrams involving one charged Higgs
boson in the loop become dominant, leaving $\lambda _{tt},\lambda _{bb}$ and 
$m_{H^{\pm }}\ $as the only free parameters in the process. However,
although Ref. \cite{Xiao} found that relatively light charged Higgs bosons
are still allowed, they also found that very heavy charged Higgs bosons are
still permitted and even required if there is a significant relative phase
between $\lambda _{tt}$ and$\ \lambda _{bb}$. Wherever a very large value of 
$m_{H^{\pm }}$ is allowed, it opens the possibility of having dominant or at
least competitive tree level diagrams even with highly suppressed values of
the couplings involving the first family, especially in the case in which at
least one of the neutral Higgs bosons is kept light. Inspired in this idea,
we shall assume in this paper that the tree level diagram is dominant.

On the other hand, in a recent previous work \cite{us}, some constraints on
LFV have been found in the framework of the two Higgs doublet model with
flavor changing neutral currents. Specifically, bounds on the vertices $\xi
_{\mu \tau },\xi _{e\tau },\xi _{\mu \mu }$,$\xi _{\tau \tau }$, were
obtained based on the $g-2$ muon factor and the leptonic decays $\mu
\rightarrow e\gamma ,\tau \rightarrow \mu \mu \mu $, $\tau \rightarrow \mu
\gamma $. Additionally, upper limits on the decays $\tau \rightarrow e\gamma 
$ and $\tau \rightarrow eee$ were estimated, finding them to be highly
suppressed respect to the present experimental sensitivity. The purpose of
this work is on one hand to complete the information about the spectrum of
the LFV matrix in the lepton sector, and on the other hand restrict some
vertices involving the first family of the quark sector, and show that such
vertices could still play a significant role in the phenomenology. Combining
bounds on the quark and lepton sector we can predict upper bounds for
leptonic decays of the $B^{0}$ mesons.

\section{Constraints in the lepton sector\label{sec:lepton}}

We shall work in the context of the two Higgs doublet model (2HDM) with
flavor changing neutral currents, the so called model type III. We shall
neglect possible relative phases between the FC vertices. The leptonic
Yukawa couplings read%
\begin{eqnarray}
-\pounds _{Y} &=&\overline{E}\left[ \frac{g}{2M_{W}}M_{E}^{diag}\right]
E\left( \cos \alpha H^{0}-\sin \alpha h^{0}\right)  \notag \\
&&+\frac{1}{\sqrt{2}}\overline{E}\xi ^{E}E\left( \sin \alpha H^{0}+\cos
\alpha h^{0}\right)  \notag \\
&&+\overline{\vartheta }\xi ^{E}P_{R}EH^{+}+\frac{i}{\sqrt{2}}\overline{E}%
\xi ^{E}\gamma _{5}EA^{0}+h.c.  \label{Yuk lepton}
\end{eqnarray}%
where $H^{0}$($h^{0}$)$\;$denote the heaviest (lightest) neutral $CP-$even
scalar, and$\;A^{0}\;$is a $CP-$odd scalar. $E\;$refers to the three charged
leptons $E\equiv \left( e,\mu ,\tau \right) ^{T}$ \ and $M_{E},\;\xi _{E}\;$%
are the mass matrix and the LFV matrix respectively, $\alpha \;$is the
mixing angle in the $CP-$even sector. We use the parameterization in which
one of the vacuum expectation values vanishes.

The decays needed to obtain our bounds are given by%
\begin{eqnarray*}
\Gamma \left( \tau ^{-}\rightarrow \mu ^{-}\mu ^{-}e^{+}\right) &=&\frac{%
m_{\tau }^{5}}{4096\pi ^{3}}\xi _{\mu \tau }^{2}\xi _{e\mu }^{2}\left[
\left( \frac{\sin ^{2}\alpha }{m_{H^{0}}^{2}}+\frac{\cos ^{2}\alpha }{%
m_{h^{0}}^{2}}-\frac{1}{m_{A^{0}}^{2}}\right) ^{2}\right. \\
&&\left. +\frac{8}{3m_{A^{0}}^{2}}\left( \frac{\sin ^{2}\alpha }{%
m_{H^{0}}^{2}}+\frac{\cos ^{2}\alpha }{m_{h^{0}}^{2}}\right) \right]
\end{eqnarray*}%
\begin{equation*}
\Gamma \left( \tau ^{-}\rightarrow \mu ^{+}\mu ^{-}e^{-}\right) =\frac{%
m_{\tau }^{5}}{6144\pi ^{3}}\xi _{\mu \tau }^{2}\xi _{e\mu }^{2}\left[
\left( \frac{\sin ^{2}\alpha }{m_{H^{0}}^{2}}+\frac{\cos ^{2}\alpha }{%
m_{h^{0}}^{2}}\right) ^{2}+\frac{1}{m_{A^{0}}^{4}}\right]
\end{equation*}%
\begin{eqnarray*}
\Gamma \left( \tau ^{-}\rightarrow \mu ^{-}e^{-}e^{+}\right) &=&\frac{%
m_{\tau }^{5}}{6144\pi ^{3}}\xi _{\mu \tau }^{2}\left\{ \left[ \sin \left(
2\alpha \right) \sqrt{\frac{G_{F}}{\sqrt{2}}}\left( \frac{1}{m_{H^{0}}^{2}}-%
\frac{1}{m_{h^{0}}^{2}}\right) m_{e}\right. \right. \\
&&\left. \left. +\xi _{ee}\left( \frac{\sin ^{2}\alpha }{m_{H^{0}}^{2}}+%
\frac{\cos ^{2}\alpha }{m_{h^{0}}^{2}}\right) \right] ^{2}+\frac{\xi
_{ee}^{2}}{m_{A^{0}}^{4}}\right\} \ .
\end{eqnarray*}%
Observe that the decays containing two identical particles in the final
state possess interferences involving the pseudoscalar Higgs boson, while
the decays with no identical leptons in the final state do not contain
interference terms with the pseudoscalar. On the other hand, in the
calculation of the decay width $\Gamma \left( \tau ^{-}\rightarrow \mu
^{+}\mu ^{-}e^{-}\right) $,$\ $we neglect diagrams containing the vertex $%
\xi _{e\tau }$ and keep only the ones proportional to $\xi _{\mu \tau }$, we
make this approximation because a previous phenomenological analysis shows a
strong hierarchy between these mixing vertices \cite{us} ($\left\vert \xi
_{e\tau }\right\vert <<\left\vert \xi _{\mu \tau }\right\vert $ by at least
five orders of magnitude).

The corresponding experimental upper limits for these rare processes are 
\cite{dataparticle}%
\begin{eqnarray}
Br \left( \tau ^{-}\rightarrow \mu ^{-}\mu ^{-}e^{+}\right) &\leq &1.5\times
10^{-6},  \notag \\
Br \left( \tau ^{-}\rightarrow \mu ^{+}\mu ^{-}e^{-}\right) &\leq &1.8\times
10^{-6},  \notag \\
Br \left( \tau ^{-}\rightarrow \mu ^{-}e^{-}e^{+}\right) &\leq &1.7\times
10^{-6}.  \label{expbounds}
\end{eqnarray}

\subsection{Bounds on $\protect\xi _{\protect\mu e}$ and $\protect\xi _{ee}$}

In a previous work \cite{us}, the LFV vertices coming from the 2HDM type
III, were constrained by using several pure leptonic processes, the
following bounds for the LFV vertices were found%
\begin{eqnarray}
\xi _{e\tau }^{2} &\lesssim &2.77\times 10^{-14}\ ,\ \left\vert \xi _{\mu
\mu }\right\vert \lesssim 1.3\times 10^{-1}\ ,  \notag \\
7.62\times 10^{-4} &\lesssim &\xi _{\mu \tau }^{2}\lesssim 4.44\times
10^{-2}\ ,  \notag \\
\left\vert \xi _{\tau \tau }\right\vert &\lesssim &2.2\times 10^{-2}\ .
\label{bounds}
\end{eqnarray}%
Such constraints are valid in most of the region of parameters. Since we
intend to complete the analysis made in \cite{us}, we shall make the same
assumptions which we summarize here for completeness. We settle $%
m_{h^{0}}\approx 115\ $GeV, and $m_{A^{0}}\gtrsim m_{h^{0}}$. In order to
cover a very wide region of parameters, we examine five cases for the
remaining free parameters of the model \cite{us}

\begin{enumerate}
\item When $m_{H^{0}}\simeq 115$ GeV.

\item When $m_{H^{0}}\simeq 300$ GeV and $\alpha =\pi /2$.

\item When $m_{H^{0}}$ is very large and $\alpha =\pi /2$.

\item When $m_{H^{0}}\simeq 300\ $GeV and $\alpha =\pi /4$.

\item When $m_{H^{0}}$ is very large and $\alpha =\pi /4$.
\end{enumerate}

\noindent For all those cases the value of the pseudoscalar mass is swept in
the range of $m_{A^{0}}\gtrsim 115$ GeV.

The vertex $\xi _{\mu e}^{2}$ can be constrained by combining the existing
limits on $\xi _{\mu \tau }^{2}$ given in Eqs. (\ref{bounds}), and
experimental upper limit on the branching ratio $Br \left( \tau
^{-}\rightarrow \mu ^{-}\mu ^{-}e^{+}\right) $ given by Eq. (\ref{expbounds}%
). Alternatively, we can constrain the same vertex from the decay $\tau
^{-}\rightarrow \mu ^{+}\mu ^{-}e^{-} $. The upper limits on $\xi _{\mu
e}^{2}$ obtained from both decays are illustrated in table (\ref{tab:mue})
for the five cases explained above. 
\begin{table}[tbh]
\centering 
\begin{tabular}{||l||l||l||}
\hline\hline
case & from $\tau ^{-}\rightarrow \mu ^{-}\mu ^{-}e^{+}$ & from $\tau
^{-}\rightarrow \mu ^{+}\mu ^{-}e^{-}$ \\ \hline\hline
1 & $\xi _{\mu e}^{2}\lesssim 5.59\times 10^{-3}$ & $\xi _{\mu
e}^{2}\lesssim 1.0\times 10^{-2}$ \\ \hline\hline
2 & $\xi _{\mu e}^{2}\lesssim 1.5\times 10^{-1}$ & $\xi _{\mu e}^{2}\lesssim
2.7\times 10^{-1}$ \\ \hline\hline
3 & unconstrained & unconstrained \\ \hline\hline
4 & $\xi _{\mu e}^{2}\lesssim 1.35\times 10^{-2}$ & $\xi _{\mu
e}^{2}\lesssim 2.43\times 10^{-2}$ \\ \hline\hline
5 & $\xi _{\mu e}^{2}\lesssim 1.67\times 10^{-2}$ & $\xi _{\mu
e}^{2}\lesssim 3.0\times 10^{-2}$ \\ \hline\hline
\end{tabular}%
\caption{{}Bounds on the mixing vertex $\protect\xi _{\protect\mu e}^{2}$,
based on the processes $\protect\tau ^{-}\rightarrow \protect\mu ^{-}\protect%
\mu ^{-}e^{+}$ and $\protect\tau ^{-}\rightarrow \protect\mu ^{+}\protect\mu %
^{-}e^{-}$ for the five cases cited in the text.}
\label{tab:mue}
\end{table}
We should observe that the upper limits obtained from $\tau ^{-}\rightarrow
\mu ^{+}\mu ^{-}e^{-}$ are less restrictive than the ones coming from $\tau
^{-}\rightarrow \mu ^{-}\mu ^{-}e^{+}$. However, both sets of constraints
lie roughly on the same order of magnitude. From table \ref{tab:mue} we can
extract a quite general bound for the vertex $\xi _{\mu e}^{2}$%
\begin{equation}
\xi _{\mu e}^{2}\leq 1.5\times 10^{-1}\ ,  \label{Xue}
\end{equation}%
valid for most of the region of parameters\footnote{%
We should bear in mind however, that none of the restrictions obtained here,
are valid for the third case explained in the text.}. It is worth saying
that other restrictions on this vertex can be gotten from $\mu \rightarrow
e\gamma $ or $\tau \rightarrow e\gamma $ assuming that only the diagrams
with a muon in the loop contribute, instead of the tau as customary.
However, bounds obtained this way are much less restrictive.

\begin{figure}[tbh]
\begin{center}
\psfrag{A}{$m_{A^0}$} \psfrag{ee}{$\xi_{ee}$} %
\includegraphics[height=5cm]{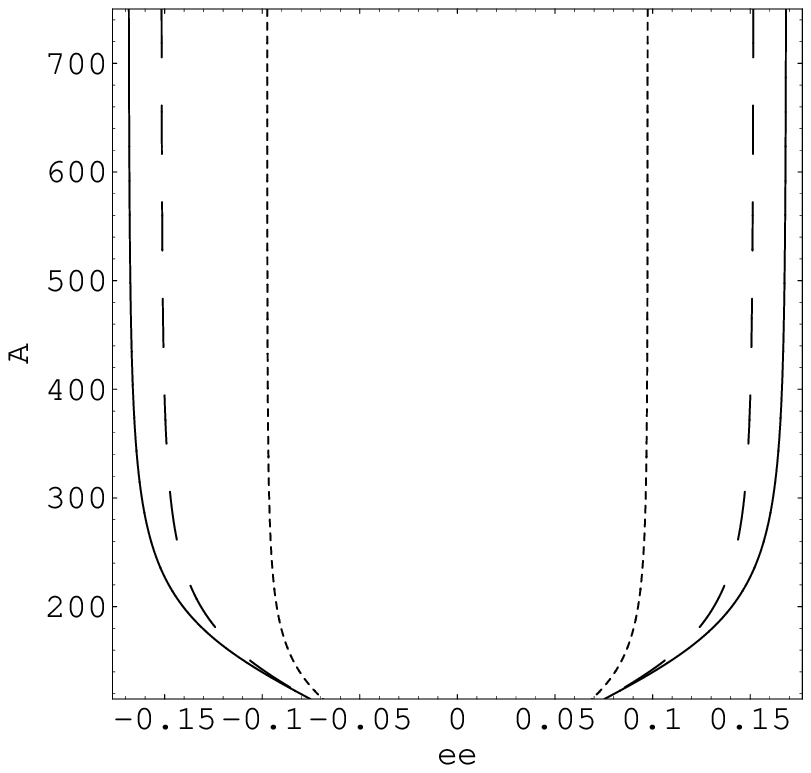} %
\includegraphics[height=5cm]{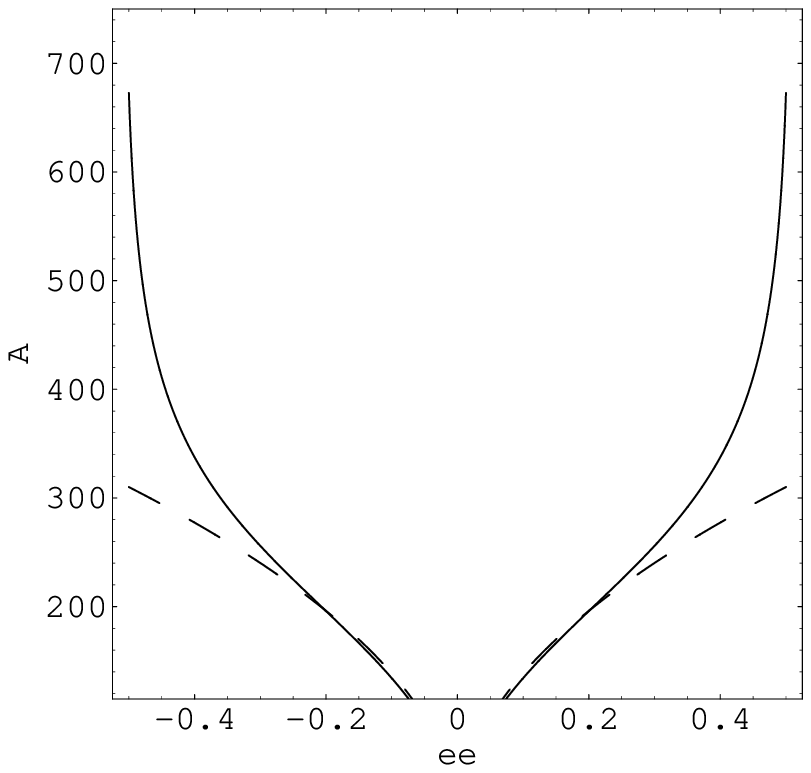}
\end{center}
\caption{Contourplots for the five cases cited in the text in the $\protect%
\xi _{ee}-m_{A^{0}\text{ }}$ plane, based on the process $\protect\tau %
^{-}\rightarrow \protect\mu ^{-}e^{+}e^{-}$. On left: Case 1 (dotted line),
case 4 (dashed line) and case 5 (solid line). On right: Case 2 (solid line)
and case 3 (dashed line).}
\label{fig:contouree}
\end{figure}

On the other hand, we can get contraints on the vertex $\xi _{ee}$ by
combining the already mentioned bounds on $\xi _{\mu \tau }^{2}$ and the
upper experimental constraints for the decay $\Gamma \left( \tau
^{-}\rightarrow \mu ^{-}e^{+}e^{-}\right) $ of Eq. (\ref{expbounds}). Since
the factor $\xi _{ee}$ cannot be factorized in contrast to the case of $\xi
_{\mu e}$, we extract its bounds in the form of contourplots in the $\xi
_{ee}-m_{A^{0}}$ plane, as shown in Fig. (\ref{fig:contouree}).
Additionally, we write in table (\ref{tab:ee}), the constraints obtained for 
$m_{A^{0}}$ very heavy and for $m_{A^{0}}\approx 115$ GeV. 
\begin{table}[tbh]
\centering 
\begin{tabular}{||l||l||l||}
\hline\hline
Case & $\left\vert \xi _{ee}\right\vert \left( m_{A^{0}}\text{ very heavy}%
\right) $ & $\left\vert \xi _{ee}\right\vert \left( m_{A^{0}}\text{ }\sim 
\text{115\ GeV}\right) $ \\ \hline\hline
1 & $\lesssim 9.75\times 10^{-2}$ & $\lesssim 6.89\times 10^{-2}$ \\ 
\hline\hline
2 & $\lesssim 5.1\times 10^{-1}$ & $\lesssim 7.41\times 10^{-2}$ \\ 
\hline\hline
3 & unconstrained & unconstrained \\ \hline\hline
4 & $\lesssim 1.5\times 10^{-1}$ & $\lesssim 7.54\times 10^{-2}$ \\ 
\hline\hline
5 & $\lesssim 1.7\times 10^{-1}$ & $\lesssim 7.53\times 10^{-2}$ \\ 
\hline\hline
\end{tabular}%
\caption{{}Bounds for the mixing matrix element $\protect\xi _{ee}$, for $%
m_{A^{0}}\simeq 115$ GeV and for $m_{A^{0}}$ very heavy. Such constraints
are based on the bounds on $\protect\xi _{\protect\mu \protect\tau }$ and
the upper limit for the decay width $\Gamma \left( \protect\tau %
^{-}\rightarrow \protect\mu ^{-}e^{+}e^{-}\right) $.}
\label{tab:ee}
\end{table}
From table (\ref{tab:ee}) we can extract general constraints for $\xi _{ee}$%
, the general bounds read%
\begin{equation*}
\left\vert \xi _{ee}\right\vert \lesssim 5.1\times 10^{-1}\ ;\ \left\vert
\xi _{ee}\right\vert \lesssim 7.54\times 10^{-2}
\end{equation*}%
for $m_{A^{0}}\approx 115$ GeV and for $m_{A^{0}}$ very heavy respectively.
We emphasize again that this prediction is valid in most of the region of
parameters but fails in the case 3 cited above, i.e. when $m_{H^{0}\text{ }}$%
is very large and $\alpha =\pi /2$.

Finally, we make a prediction about the upper limit for the branching ratio
of the process$\ \tau ^{-}\rightarrow \mu ^{+}e^{-}e^{-}$, based on the
limits on $\xi _{e\mu }$ shown in table \ref{tab:mue} and the limits on $\xi
_{e\tau }$ shown in Eqs. (\ref{bounds}), the results are collected in table %
\ref{tab:taeemu}. We see that the upper limits shown in table (\ref%
{tab:taeemu}) are at least ten orders of magnitude smaller than the present
experimental upper limit $Br \left( \tau ^{-}\rightarrow \mu
^{+}e^{-}e^{-}\right) \leq 1.5\times 10^{-6}\ $GeV, (except for the third
case). In addition, table \ref{tab:pred} collects the results for the
general upper limits of three leptonic decays involving the vertex $\xi
_{e\tau }$.\ The strong suppression of these processes might be anticipated
because of its proportionality to $\xi _{e\tau }^{2}$ which is much more
restricted than the others \cite{us}.

\begin{table}[tbh]
\centering 
\begin{tabular}{||c||c||}
\hline\hline
Case & $Br (\tau ^{-}\rightarrow \mu ^{+}e^{-}e^{-})$ \\ \hline\hline
1 & $\lesssim 9.5\times 10^{-18}$ \\ \hline\hline
2 & $\lesssim 3.2\times 10^{-17}$ \\ \hline\hline
3 & Unconstrained \\ \hline\hline
4 & $\lesssim 1.3\times 10^{-17}$ \\ \hline\hline
5 & $\lesssim 1.4\times 10^{-17}$ \\ \hline\hline
\end{tabular}%
\caption{{}Upper limits for the branching ratio $Br \left( \protect\tau %
^{-}\rightarrow e^{-}e^{-}\protect\mu ^{+}\right) $, based on the contraints
obtained for the LFV vertices $\protect\xi _{\protect\mu e}$ and $\protect%
\xi _{e\protect\tau }$. The experimental upper limit is $1.5\times 10^{-6}$}
\label{tab:taeemu}
\end{table}

\begin{table}[tbp!]
\centering 
\begin{tabular}{||l||l||}
\hline\hline
Predictions & Experim. limits \\ \hline\hline
$Br \left( \tau ^{-}\rightarrow e^{-}\gamma \right) \lesssim 6.6\times
10^{-16}$ & $2.7\times 10^{-6}$ \\ \hline\hline
$Br \left( \tau ^{-}\rightarrow e^{+}e^{-}e^{-}\right) \lesssim 2.2\times
10^{-17}$ & $2.9\times 10^{-6}$ \\ \hline\hline
$Br\left( \tau \rightarrow \mu ^{+}e^{-}e^{-}\right) \lesssim 3.2\times
10^{-17}$ & $1.5\times 10^{-6}$ \\ \hline\hline
\end{tabular}%
\caption{{}Upper limits predicted for some lepton decays. All of them are
highly suppressed respect to the current experimental upper limit.}
\label{tab:pred}
\end{table}

\section{Constraints in the quark sector\label{sec:quark}}

We shall obtain constraints on the quark sector by using the experimental
information from $B_{d,s}^{0}$ lepton decays and $\Delta m_{B_{d}^{0}}$. The 
$B_{d}^{0}$ measurements are dominated by the asymmetric $B$ factories \cite%
{Belle}, while the $B_{s}^{0}$ measurements come from hadron colliders \cite%
{hadroncol}. At the tree level, the decays $B_{d}^{0}\rightarrow ll^{\prime
} $ depend on the product $\xi _{ll^{\prime }}^{2}\xi _{db}^{2}$ and the
pseudoscalar Higgs boson mass only. In the framework of the 2HDM-III they are%
\begin{eqnarray}
\Gamma \left( B_{q}^{0}\rightarrow l^{-}l^{\prime +}\right) &=&\frac{%
m_{B_{q}}f_{B_{q}}^{2}\xi _{qb}^{2}\xi _{ll^{\prime }}^{2}}{32\pi \left(
m_{b}+m_{q}\right) ^{2}m_{A^{0}}^{4}}\left[ 2(m_{B_{q}}^{2}-m_{l}^{2}-m_{l^{%
\prime }}^{2})-4m_{l}m_{l^{\prime }}\right] \times  \notag \\
&&\sqrt{\left[ m_{B_{q}}^{2}-\left( m_{l^{\prime }}+m_{l}\right) ^{2}\right] %
\left[ m_{B_{q}}^{2}-\left( m_{l}-m_{l^{\prime }}\right) ^{2}\right] }.
\label{Bdecth}
\end{eqnarray}%
Where $f_{B_{q}}$ represent the $B_{d}^{0}$ meson decay constant whose value
has been taken from \cite{decconst}. The present experimental upper bounds
for these decays at 90\% C.L. are \cite{dataparticle}%
\begin{eqnarray*}
Br \left( B_{d}^{0}\rightarrow e^{-}\mu ^{+}\right) &\leq &1.5\times 10^{-6}
\\
Br \left( B_{d}^{0}\rightarrow e^{-}\tau ^{+}\right) &\leq &5.3\times 10^{-4}
\\
Br \left( B_{d}^{0}\rightarrow \mu ^{-}\tau ^{+}\right) &\leq &8.3\times
10^{-4} \\
Br \left( B_{s}^{0}\rightarrow e^{-}\mu ^{+}\right) &\leq &6.1\times 10^{-6}
\end{eqnarray*}%
Consequently, we can get upper limits for this products of mixing vertices
by using the upper bound for these decays. In particular, from $%
B_{d}\rightarrow \mu \tau $ we find%
\begin{equation*}
\xi _{\mu \tau }^{2}\xi _{db}^{2}\lesssim \left( 5.45\times
10^{-15}GeV^{-4}\right) m_{A^{0}}^{4}\ .
\end{equation*}%
On the other hand, since we have a lower bound on the mixing vertex $\xi
_{\mu \tau }^{2}$ we can obtain an upper bound for $\xi _{db}^{2}$ alone%
\begin{equation}
\xi _{db}^{2}\lesssim \left( 7.15\times 10^{-12}GeV^{-4}\right)
m_{A^{0}}^{4}\ .  \label{Xdb}
\end{equation}%
We can obtain a similar bound for the product $\xi _{e\tau }^{2}\xi
_{db}^{2} $ based on the upper limit for $B_{d}\rightarrow e\tau $.
Nevertheless, a better bound for this product is obtained by combining Eqs. (%
\ref{bounds}), (\ref{Xdb}). The same situation occurs for the product $\xi
_{e\mu }^{2}\xi _{db}^{2}$, since the present bounds on the decays $%
B_{d}\rightarrow e\mu $ cannot provide a better constraint for this product
that the bound obtained from Eqs. (\ref{Xue}), (\ref{Xdb}).

Furthermore, from the upper limit for $Br \left( B_{s}^{0}\rightarrow e\mu
\right) $ we find an upper bound for the product $\xi _{sb}^{2}\xi _{e\mu
}^{2}$ 
\begin{equation*}
\xi _{sb}^{2}\xi _{e\mu }^{2}\leq 2.38\times 10^{-17}m_{A}^{4}
\end{equation*}

As we see from Eq. (\ref{Bdecth}) and the bounds in this section, the latter
blow up rapidly when $m_{A^{0}}$ grows. Indeed, for heavy values of the
pseudoscalar mass, the one loop contributions could be sizeable \cite{Wolf}
introducing more free parameters to the model. Nevertheless, if we consider
a quite light pseudoscalar i.e. a mass no far from the electroweak scale,
the tree level contribution is dominant and the bounds above are quite
restrictive. Table \ref{tab:quark1} shows some typical values for the upper
limits above for $115\ $GeV$\lesssim m_{A^{0}}\lesssim 250$\ GeV.

\begin{table}[h]
\label{tabla3}
\par
\begin{center}
\begin{tabular}{||c||c||c||c||}
\hline\hline
$m_{A}$(GeV) & 115 & 200 & 250 \\ \hline\hline
$\xi _{db}^{2}(\times 10^{-3})$ & $1.25$ & $11.44$ & $27.92$ \\ \hline\hline
$\xi _{db}^{2}\xi _{e\tau }^{2}(\times 10^{-17})$ & $3.46$ & $31.69$ & $%
77.35 $ \\ \hline\hline
$\xi _{db}^{2}\xi _{e\mu }^{2}(\times 10^{-4})$ & 1.87 & 17.16 & 41.88 \\ 
\hline\hline
$\xi _{sb}^{2}\xi _{e\mu }^{2}(\times 10^{-8})$ & 0.42 & 3.82 & 9.32 \\ 
\hline\hline
\end{tabular}%
\end{center}
\caption{Bounds for $\protect\xi _{db}^{2}$, $\protect\xi _{sb}^{2}$ and
their products with leptonic vertices for pseudoscalar bosons lying roughly
in the electroweak scale.}
\label{tab:quark1}
\end{table}

\subsection{Constraints from $B_{d}^{0}-\overline{B}_{d}^{0}$ mixing\label%
{sec:BB}}

Now let us use the $\Delta M_{B_{d}^{0}}$ parameter to constrain the vertex $%
\xi _{db}$ involving the first family. The invariant amplitude at the tree
level for $B_{q}^{0}-\bar{B}_{q}^{0}$ mixing in our model is given by

\begin{eqnarray}
\langle B_{q}^{0}|H_{W}|\overline{B}_{q}^{0}\rangle &=&-\frac{2}{m_{H}^{2}}%
R_{qbH}^{2}\langle B_{q}^{0}|\bar{b}q\bar{b}q|\overline{B}_{q}^{0}\rangle -%
\frac{2}{m_{h}^{2}}R_{qbh}^{2}\langle B_{q}^{0}|\bar{b}q\bar{b}q|\overline{B}%
_{q}^{0}\rangle  \notag \\
&&-\frac{2}{m_{A}^{2}}\bar{R}_{qbA}^{2}\langle B_{q}^{0}|\overline{b}\gamma
_{5}q\overline{b}\gamma _{5}q|\overline{B}_{q}^{0}\rangle ,  \label{deltamH}
\end{eqnarray}%
where $R_{qbh}$ are the coefficients of the Feynman rules with $q=d,s$.\ In
terms of the operators $O_{\Delta F=2}^{F}$ defined in \cite{Reina, Bruce},
we have that 
\begin{equation}
\langle B_{q}^{0}|O^{B_{q}^{0}}|\overline{B}_{q}^{0}\rangle =B_{B}\langle 
\overline{B}_{q}^{0}|O^{B_{q}^{0}}|\overline{B}_{q}^{0}\rangle _{VIA},
\end{equation}%
where $VIA$ denotes the Vacuum Insertion Approximation and $B_{B_{q}}$ is
the vacuum saturation coefficient. The operators that we need in our case
are 
\begin{equation}
O_{S}^{B_{q}^{0}}=(\bar{b}q)(\bar{b}q)\ \ ;\ \ O_{P}^{B_{q}^{0}}=(\bar{b}%
\gamma _{5}q)(\bar{b}\gamma _{5}q).
\end{equation}%
In addition, based on the expressions shown in section VI of the first of
Refs. \cite{Reina}, we introduce the factors $M_{S}^{B}$ and $M_{P}^{B}$ in
terms of the $\Delta F=2\ $matrix elements of the only two operators which
do not vanish in the vacuum 
\begin{eqnarray}
M_{S}^{B_{q}^{0}} &=&\langle B^{0}|O_{S}^{B_{q}^{0}}|\overline{B}^{0}\rangle
_{VIA}=-\frac{1}{6}M_{P}^{0,B_{q}^{0}}+\frac{1}{6}M_{A}^{0,B_{q}^{0}}  \notag
\\
M_{P}^{B} &=&\langle B^{0}|O_{P}^{B^{0}}|\overline{B}^{0}\rangle _{VIA}=%
\frac{11}{6}M_{P}^{0,B}-\frac{1}{6}M_{A}^{0,B},
\end{eqnarray}%
where 
\begin{eqnarray}
M_{P}^{0,B_{q}^{0}} &=&\langle B_{q}^{0}|\overline{\psi }_{B_{q}^{0}}\gamma
_{5}\psi _{q}|0\rangle \langle 0|\overline{\psi }_{B_{q}^{0}}\gamma _{5}\psi
_{q}|\overline{B}_{q}^{0}\rangle =-f_{B_{q}^{0}}^{2}\frac{m_{B_{q}^{0}}^{4}}{%
(m_{b}+m_{q})^{2}}  \notag \\
M_{A}^{0,B_{q}^{0}} &=&\langle B_{q}^{0}|\overline{\psi }_{B_{q}^{0}}\gamma
_{\mu }\gamma _{5}\psi _{q}|0\rangle \langle 0|\overline{\psi }%
_{B_{q}^{0}}\gamma _{\mu }\gamma _{5}\psi _{q}|\overline{B}_{q}^{0}\rangle
=f_{B_{q}^{0}}^{2}m_{B_{q}^{0}}^{2}.  \label{mAB}
\end{eqnarray}%
Using Eqs. (\ref{deltamH})-(\ref{mAB}), we find that the contribution from
new physics to the parameter $\Delta M_{B_{q}^{0}}$ reads 
\begin{eqnarray}
\Delta M_{NP} &=&-2\Re \langle B_{q}^{0}|H_{W}|\overline{B}_{q}^{0}\rangle 
\notag \\
&=&f_{B_{q}^{0}}^{2}m_{B_{q}^{0}}^{2}\xi _{qb}^{2}\Big[\frac{1}{3m_{h}^{2}}%
\cos ^{2}\alpha \left( \frac{m_{B_{q}^{0}}^{2}}{(m_{q}+m_{b})^{2}}+1\right) 
\notag \\
&&+\frac{1}{3m_{H}^{2}}\sin ^{2}\alpha \left( \frac{m_{B_{q}^{0}}^{2}}{%
(m_{q}+m_{b})^{2}}+1\right) +\frac{1}{3m_{A}^{2}}\left( 11\frac{%
m_{B_{q}^{0}}^{2}}{(m_{q}+m_{b})^{2}}+1\right) \Big].  \label{dMNP}
\end{eqnarray}

We shall estimate bounds for $\xi _{db}$ by using $\Delta M_{B_{d}^{0}}$
coming from $B_{d}^{0}-\overline{B}_{d}^{0}$ mixing. The predictions for the
standard model (SM) $\Delta M_{SM}$ and the experimental value $\Delta
M_{EXP}\ $have been taken from \cite{Xiao} by using symmetrical
uncertainties for the sake of simplicity%
\begin{eqnarray}
\Delta M_{SM} &=&0.506\pm 0.198ps^{-1}  \notag \\
\Delta M_{EXP} &=&0.502\pm 0.007ps^{-1}  \label{dmmeasurement}
\end{eqnarray}%
The maximum room for the new physics reads%
\begin{equation}
\Delta M_{NP}\leq \Delta M_{EXP}^{0}-\Delta M_{SM}^{0}-\sqrt{%
E_{SM}^{2}+E_{EXP}^{2}}
\end{equation}%
where $\Delta M_{EXP}^{0}$ and $\Delta M_{SM}^{0}$ represent the central
values of $\Delta M_{EXP}$ and $\Delta M_{SM}$ respectively. Furthermore, $%
E_{SM}$ and $E_{EXP}$ are the uncertainties associated to the standard model
and experimental estimations respectively. All of them are given by Eq. (\ref%
{dmmeasurement}). So we can constrain $\xi _{db}$ based on the values of $%
\Delta M_{SM}$ and $\Delta M_{EXP}$ for the $B_{d}^{0}$ meson. We found the
results shown in table \ref{tab:bd} for the five cases, when we consider $%
m_{A}=115$ GeV and $m_{A^{0}}$ very large. 
\begin{table}[tbp]
\label{tabf}
\par
\begin{center}
\begin{tabular}{||c||c||c||}
\hline\hline
Case & $|\xi _{db}|$ ($m_{A}=115$ GeV ) & $|\xi _{db}|$( $m_{A}$ very large )
\\ \hline\hline
1 & $2.69\times 10^{-6}$ & $7.37\times 10^{-6}$ \\ \hline\hline
2 & $2.86\times 10^{-6}$ & $19.22\times 10^{-6}$ \\ \hline\hline
3 & $2.89\times 10^{-6}$ & unconstrained \\ \hline\hline
4 & $2.77\times 10^{-6}$ & $9.73\times 10^{-6}$ \\ \hline\hline
5 & $2.79\times 10^{-6}$ & $10.42\times 10^{-6}$ \\ \hline\hline
\end{tabular}%
\end{center}
\caption{Constraints on $|\protect\xi _{db}|$ for $m_{A}=115$ GeV and $m_{A}$
very large, based on the data of $\Delta M_{B_{d}}$ }
\label{tab:bd}
\end{table}

Now, using the bounds on $\xi _{db}^{2}$ obtained from the $\Delta
M_{B_{d}^{0}}$ and combining them with the allowed values for the vertex $%
\xi _{\mu \mu }$ in Eq. (\ref{bounds}), we shall predict the maximum
contribution of this new physics to the decay $B_{d}^{0}\rightarrow \mu 
\overline{\mu }$. Such decay has already been considered in the literature
in the framework of the two Higss doublet model with and without FCNC \cite%
{Savage}. The predicted upper bounds for this decay for the five cases
explained in the text, are displayed in table \ref{tab:Bmumu}
\begin{table}[tbp]
\begin{center}
\begin{tabular}{||c||c||c||}
\hline\hline
Case & $Br (B_{d}\rightarrow \mu \overline{\mu })$ ($m_{A}=115$ GeV ) & $Br
(B_{d}\rightarrow \mu \overline{\mu })\ $( $m_{A}$ very large ) \\
\hline\hline
1 & $2.14\times 10^{-8}$ & $1\times 10^{-8}$ \\ \hline\hline
2 & $2.3\times 10^{-8}$ & $1\times 10^{-8}$ \\ \hline\hline
3 & $2.32\times 10^{-8}$ & unconstrained \\ \hline\hline
4 & $2.21\times 10^{-8}$ & $1\times 10^{-8}$ \\ \hline\hline
5 & $2.23\times 10^{-8}$ & $1\times 10^{-8}$ \\ \hline\hline
\end{tabular}%
\end{center}
\caption{ Upper limits for the branching ratio $Br\left( B_{d}\rightarrow
\protect\mu \overline{\protect\mu }\right) $ based on the upper limits for $%
\protect\xi _{db}$ and the allowed values for $\protect\xi _{\protect\mu 
\protect\mu }$.}
\label{tab:Bmumu}
\end{table}
and the SM prediction which was calculated by avoiding the big uncertainties
of $f_{B_{d}^{0}}$ is given by \cite{Buras}%
\begin{equation*}
Br\left( B_{d}\rightarrow \mu \overline{\mu }\right) _{SM}=1\times 10^{-10}
\end{equation*}%
Taking into account the upper limits of $\left\vert \xi _{db}\right\vert \ $ we obtain that the tree level contribution to
this process coming from the 2HDM can be comparable and even dominant
respect to the SM contribution. It shows that, although the mixing vertices
involving the first generation are highly suppressed, is still possible for
them to play a role in the phenomenology.

\section{Conclusions\label{sec:conclusions}}

We have found constraints on the whole spectrum of the mixing matrix of
leptons, by using purely leptonic processes. A strong hierarchy between the
vertices $\xi _{\mu \tau }$ and $\xi _{e\tau }$ is manifest. Effectively, as
it is shown in table (\ref{tab:pred}), decays involving the $\xi _{e\tau }$
vertex, are highly suppressed respect to the current instrumental
sensitivity. The constraints on the rest of LFV vertices are much milder.
However, current prospects to increase the statistics concerning LFV decays
could improve such bounds significantly.

In addition, we constrain some FC couplings in the quark sector by using
experimental limits on leptonic $B_{d,s}^{0}$ decays as well as the $%
B_{d}^{0}-\overline{B}_{d}^{0}$ mixing. The leptonic $B$ decays provide
constraints on the products of lepton and quark FC couplings. On the other
hand, by assuming that the charged Higgs boson is sufficiently heavy, the $%
B_{d}^{0}-\overline{B}_{d}^{0}$ mixing can be used to constrain the vertex $%
\xi _{db}$. We point out that although the $B_{d}^{0}-\overline{B}_{d}^{0}$
mixing imposes severe restrictions to this vertex, the upper limits for $\xi
_{db}\ $could still give a sizeable and even dominant contribution to the
decay $B_{d}^{0}\rightarrow \mu \overline{\mu }$ respect to the SM
contribution. Consequently, this vertex of the first generation can still be
important for phenomenological calculations.

The authors acknowledge the financial support by Fundaci\'{o}n Banco de la
Rep\'{u}blica, Colciencias, and DINAIN.

\end{document}